\documentclass[11pt,letterpaper,twocolumn]{article}

\usepackage[utf8]{inputenc}
\usepackage[english]{babel}
\usepackage{float}
\usepackage{xcolor}
\usepackage{verbatim}
\usepackage{mwe}
\usepackage{charter}
\usepackage{afterpage}
\usepackage{amsmath}
\usepackage{appendix}
\usepackage{ragged2e}
\usepackage{array}
\usepackage{etoolbox}
\usepackage{hyperref}
\usepackage{fancyhdr}
\usepackage{booktabs}
\usepackage{arydshln}
\usepackage[justification=justified,singlelinecheck=false,labelfont=bf,format=plain]{caption}
\usepackage[justification=justified,singlelinecheck=false,labelfont=bf,format=plain]{subcaption}
\usepackage{enumitem}
\usepackage[bottom=2.5cm,top=2.0cm,left=1.5cm,right=1.5cm]{geometry}
\usepackage{graphicx}
\usepackage{indentfirst}
\usepackage{mathtools}
\usepackage{multirow}
\usepackage{pdfpages}

\usepackage{subfiles}
\usepackage[compact]{titlesec}
\usepackage{blindtext}
\usepackage{stfloats}
\usepackage{lipsum}

\newcolumntype{L}[1]{>{\raggedright\let\newline\\\arraybackslash\hspace{0pt}}m{#1}}
\newcolumntype{C}[1]{>{\centering\let\newline\\\arraybackslash\hspace{0pt}}m{#1}}
\newcolumntype{R}[1]{>{\raggedleft\let\newline\\\arraybackslash\hspace{0pt}}m{#1}}

    \setlist[itemize,1]{label=$\bullet$}
    \setlist[itemize,2]{label=$\circ$}
    \setlist[itemize,3]{label=$-$}
    \setlist{nosep}

\titlelabel{\thetitle.\quad}

\makeatletter
\patchcmd{\headrule}{\hrule}{\color{black}\hrule}{}{} % headrule
\patchcmd{\footrule}{\hrule}{\color{black}\hrule}{}{} % footrule
\makeatother

\definecolor{blueM}{cmyk}{1.0,0.49,0.0,0.47}

\begin{document}
\twocolumn[\begin{@twocolumnfalse}

\hspace{25pt}
\begin{minipage}{0.9\textwidth}
\vspace{5mm}
    \Large{\textbf{Molecular Rotors as intracellular probes of Red Blood Cell stiffness}} %
    \vspace{3mm}

    \large{Alice Briole$^a$, Thomas Podgorski$^b$, Bérengère Abou$^a$} % Si solo hay un autor borrar el apartado sin borrar el ultimo }
 \newline
    %Si solo hay un asesor borrar el ``1'' y el apartado de Asesor 2
    \fontsize{0.35cm}{0.5cm}\selectfont \textit{$^{a}$~Laboratoire Matière et Systèmes Complexes, UMR 7057 CNRS - Université de Paris, 75013 Paris, France; E-mail: berengere.abou@univ-paris-diderot.fr}
    \newline 
  \textit{$^{b}$~Laboratoire Rhéologie et Procédés, UMR 5520  CNRS-UGA-G.INP - Domaine Universitaire - BP 53
38041 Grenoble Cedex 9, France; E-mail: thomas.podgorski@univ-grenoble-alpes.fr}
    \vspace{1mm} 
    
    %\today % FECHA

\end{minipage}

\small

\vspace{11pt}

\centerline{\rule{0.95\textwidth}{0.4pt}}

\begin{center}
    
    \begin{minipage}{0.9\textwidth}
        % RESUMEN
        \noindent The deformability of red blood cells is an essential parameter that 
controls the rheology of blood as well as its circulation in the body. Characterizing the rigidity of the cells and their heterogeneity in a blood sample is thus a key point in the understanding of occlusive phenomena, particularly in the case of erythrocytic diseases in which healthy cells coexist with pathological cells. However, measuring intracellular rheology in small biological compartments requires the development of specific techniques. We propose a technique based on molecular rotors -- viscosity-sensitive fluorescent probes -- to evaluate the above key point. DASPI molecular rotor has been identified with spectral fluorescence properties decoupled from those of hemoglobin, the main component of the cytosol. After validation of the rotor as a viscosity probe in model fluids, we showed by confocal microscopy that, in addition to binding to the membrane, the rotor penetrates spontaneously and uniformly into red blood cells. Experiments conducted on temperature-stiffened red blood cells show that molecular rotors can probe their overall rigidity. A simple model allowed us to separate the contribution of the cytosol from that of the membrane, providing a quantification of cytosol rigidification with temperature, consistent with independent measurements of the viscosity of hemoglobin solutions. Our experiments demonstrate that the rotor can be used to quantify the intracellular rheology of red blood cells at the cellular level, as well as the heterogeneity of this stiffness in a blood sample. This technique opens up new possibilities for biomedical applications, diagnosis and disease monitoring.
    \end{minipage}
\end{center}
\centerline{\rule{0.95\textwidth}{0.4pt}}
\vspace{15pt}
\end{@twocolumnfalse}]

\justify

\section{Introduction}

Red blood cells (RBCs) are the main cellular component of blood and perform the essential function of carrying oxygen through the body. These cells essentially consist of a deformable membrane containing a hemoglobin solution considered as a Newtonian fluid \cite{cokelet_1968} in the healthy case where the mean corpuscular hemoglobin concentration (MHC) is approximately $32$ g/dL. The physical properties of the components -- membrane bending and shear resistance, cytosol viscosity -- condition the deformability of the red blood cells and their flow dynamics. This is crucial for the proper functioning of the circulatory system, particularly for microcirculation where RBCs have to squeeze through complex networks of narrow capillaries.

The physical properties of red blood cells vary over their lifetime ($\sim 100-120$ days). Cells become denser due to dehydration, resulting in a $25-30\%$ increase in the concentration of corpuscular hemoglobin between the more and less dense fractions \cite{bosch1992,bosch1994}. Since the viscosity of hemoglobin solutions varies sharply between concentrations of $30$ and $40$ g/dL \cite{ross1977,chien1970}, the densification of red blood cells leads to an increase in their internal viscosity from about $5$ to $20$ mPa.s, resulting in significant heterogeneity of blood samples \cite{waugh1992,franco2013}. 
This is not without consequences. It is well documented in the literature on erythrocyte suspension flows and rheology that the dynamics of cells (or more generally capsules and vesicles) in simple and complex flows strongly depends on the viscosity ratio between the internal fluid (cytosol) and the suspending fluid (plasma or buffer) \cite{dupire2012, fischer2013, mauer2018, minetti2019, losserand2019, shen2016}. In addition, the reduced deformability of senescent cells affects their elimination by mechanical filtration through the spleen.  
Several pathological conditions also lead to alterations in the deformability of red blood cells. This is the case of sickle cell disease, where red blood cells become rigid under the effect of deoxygenation, with serious consequences on the microcirculation in small capillaries, which can lead to occlusion phenomena and vaso-occlusive crises.

All this indicates that the deformability and stiffness of red blood cells is a key haemorheological parameter for diagnosis \cite{banerjee1998, du2015}. 

It is likely that jamming and occlusion phenomena are strongly influenced by the presence and quantity of rigid cells rather than by the average rigidity of the sample: a small fraction of rigid red blood cells (i.e. irreversibly sickled cells in the case of sickle cell disease) may be sufficient to nucleate occlusions despite an apparently low average rigidity value.
The details of the distribution of mechanical properties of RBCs -- which highlights the heterogeneity of a given blood sample -- is therefore an information that could be useful in assessing the risk of occlusion.

A number of techniques have been proposed to quantify the deformability of red blood cells. 
Atomic force microscopy (AFM) has been used to characterize the stiffness of RBCs at the cellular level, in healthy and pathological cases \cite{maciaszek2011a,maciaszek2011b}. The technique can reveal significant variations in cell elasticity in a research context, but requires %skilled experimentalists and
heavy equipment, and is not suitable for statistical measurements of large-scale samples in a clinical practice.
Ektacytometry, developed in the 1980’s, is based on light diffraction through a blood sample in shear flow, to measure a RBC elongation index\cite{bessis1980}, with the intrinsic limitation of providing only the average deformability of the red blood cells in the sample and 

challenging interpretation of the results\cite{rabai2014,nikitin2015,sosa2014,renoux2015}. 
 
In recent years, microfluidic techniques have been proposed to characterize deformability at the cellular level \cite{guo2014, kim2015}. 

Although they partly answer the question of the dispersion of RBC properties in a sample, they require heavy image processing and shape analysis giving indirect information on the mechanical properties. 
Other techniques have focused specifically on membrane shear and bending moduli, including pipette aspiration\cite{discher1994}, electric field deformation\cite{Engelhardt1984}, optical tweezers\cite{sleep1999,dao2003}, and optical microrheology\cite{Puig-de-Morales2007,amin2007,wang2011}.
It appears that there is currently a need for sub-cellular rheology techniques allowing quantitative measurements of the mechanical properties of different parts of the cell and operating on large samples. As further evidence, in the particular case of RBCs, measurements of the viscosity of the internal hemoglobin solution have mostly been performed by extracting the cell's hemoglobin content by lysis followed by conventional rheology techniques on relatively large samples\cite{chien1970, monkos1994, iino1997}. 

Molecular rotors (MRs) have recently proven their usefulness in measuring the rheological properties of intracellular media for the purpose of characterization and
diagnostics \cite{kuimova2008,kuimova2009,woodcock2019}. 
This is a group of fluorescent molecules having the ability to form twisted intra-molecular charge transfer (TICT) states, upon photoexcitation.
Relaxation from the TICT conformation occurs either by emission of a red-shifted emission band or by non-radiative relaxation, with a lower excited-state/ground-state energy gap than the planar, locally excited (LE) state \cite{haidekker2010a}. For viscosity measurements, molecular rotors with a single emission band are most preferred because emission from the planar (non-twisted) LE conformation is highly sensitive towards viscosity\cite{loufty1982,law1980,haidekker2010a}, and relatively insensitive towards the solvent polarity \cite{allen2005,kung1986,haidekker2005}. 
It has been demonstrated in a wide range of viscosities and in both polar and non polar fluids, that the
quantum yield $\phi$ of the LE peak increases with viscosity $\eta$, according the so-called Förster-Hoffmann equation, $\log \phi= C + b \log \eta$, where $C$ and $b$ are solvent and dye dependent constants\cite{loufty1982, forster1971,law1980,iwaki1993,haidekker2005,wang2009}. These rheology dependent properties have been used for real-time monitoring of aggregation and
polymerization processes \cite{loufty1983}, to report
aggregation and protein conformational changes \cite{hawe2010,kung1989}, and study phospholipid bilayers and cell membranes \cite{kung1986, furuno1992, viriot1998}.

Here we study the ability of a selected molecular rotor to quantify the heterogeneity of a blood sample, and to characterize the rigidity and intracellular rheology of red blood cells.
After characterizing its spectral properties, which should not interfere with those of hemoglobin, and
its effectiveness as a viscosity probe in simple fluids, we demonstrate that, in addition to attaching to the membrane, the rotor penetrates spontaneously and uniformly into the red blood cells.
We show that it is possible to distinguish red blood cells with different levels of deformability in a blood sample, and to quantify the heterogeneity of this sample. The evolution of the average stiffness of a blood sample and the distribution of stiffness in the sample are measured by varying the temperature. The fluorescence of the cytosol alone is estimated at different temperatures using a simple geometrical model of RBC allowing the separation of cytosol and membrane contributions. The estimated rigidification of cytosol with temperature is consistent with
independent measurements of the viscosity of hemoglobin solutions. This opens up opportunities for the development of rapid and inexpensive hematological tests for diagnosis and clinical monitoring of patients.

\section{Materials and Methods}

\subsection{Red Blood Cell samples and hemoglobin solutions}
\label{prep}
{\bf Preparation and stiffening of red blood cells}. 
Blood samples from healthy donors were provided by Etablissement Français du Sang (EFS Rhône-Alpes and EFS Île-de-France). Red Blood Cells were washed three times by dilution in phosphate buffer saline (PBS) and centrifugation at $1,000 \times g$. They were resuspended in PBS to reach a hematocrit 
%concentration 
value of $0.25 \%$. A $40$ mM stock solution of DASPI (trans-4-[4-(Dimethylamino)styryl]-1-methylpyridinium iodide, ref. 336408 from Sigma-Aldrich) in DMSO was prepared and then diluted in PBS to form two working solutions with concentrations of $1$ mM and $10$ mM. The above RBC suspensions were stained by adding $10 \%$ (v/v) of the DASPI solution to reach a final DASPI concentration of $0.1$ mM or $1$ mM. They were incubated for $2$ hours at room temperature to allow DASPI to diffuse through the membranes, into the cytosol.
Modulation of RBC mechanical properties was induced by temperature variations. RBCs were studied at $18^\circ$C, $24^\circ$C, $30^\circ$C, $37^\circ$C and $43^\circ$C.

\noindent {\bf Hemoglobin solutions}. 
Hemoglobin solutions were prepared from blood samples. Red blood cells were washed three times in PBS buffer.
After a final centrifugation at $1,200 \times g$ for $10$ min, all supernatant was removed. The RBC pellet was exposed to $3$ freeze-thaw cycles ($-80^\circ$C to $37^\circ$C) to lyse red blood cell membranes. Membrane residues were then extracted by addition of CCl$_4$ to the lysate ($1/3$ of the volume), strong agitation and ultra-centrifugation at $10,000 \times g$ for $15$ min.
Hemoglobin solutions of various concentrations were obtained by dilution in PBS or by concentration using centrifugal ultrafiltration units (Vivaspin 2 PES 30 kDa from Sartorius). The hemoglobin concentration of the final solutions was measured by colorimetry (MAK 115 Hemoglobin Assay Kit from Sigma-Aldrich). The pH of hemoglobin solutions 
%was measured and 
was of constant value $7\pm0.2$ for all concentrations.

\noindent {\bf Preparation of model fluids for viscosity measurements}. 
Water/glycerol and glycerol/ethylene glycol mixtures were prepared at different glycerol concentrations ($0\%$ v/v to $70\%$ v/v). DASPI ($8$ mM stock solution in PBS) was added to reach a final concentration of $0.01$ mM, $0.1$ mM or $1$ mM.

\subsection{Sample temperature control}\label{Tcontrol}

For the microrheology and fluorescence microscopy experiments (sections \ref{sec:FH} and \ref{sec:stiffening}), the samples were observed on an inverted Leica DMI 8 microscope. The temperature of the samples was controlled and varied by an objective heater through the immersion oil in contact with the sample (Bioptechs Inc., Butler, PA, USA) with an accuracy of $\pm 0.1^\circ$C.

\subsection{Particle tracking microrheology}
The viscosity of the solutions (water/glycerol, glycerol/ethylene glycol and hemoglobin) was measured by particle tracking microrheology. In experiments described in section \ref{sec:FH}, the viscosity of the solutions was measured in situ with microrheology experiments, before proceeding with the fluorescence measurements. The technique is based on the Brownian motion of
microscopic probe particles immersed in the fluid. Microrheology experiments are generally carried out on small volumes of about $1$ $\mu$L.
The technique is interesting in biological samples where minute volumes can be studied\cite{Abou2010-JRSI,Abou-jeb2019}, and also in heterogeneous samples where the technique gives access to the amplitude of the heterogeneities in the sample\cite{Abou2011-SM}. 
Calibrated polystyrene beads ($0.994$ $\mu$m in diameter) were dispersed in the solutions to be studied at a volume fraction of less than $1\%$. The thermal motion of the tracers was recorded for $20$ s at a rate of $100$ Hz with an OrcaFlash v2+ fast sCMOS camera. This camera is mounted on a Leica inverted bright field microscope DMI 8, including an oil immersion objective with 100X magnification ($NA=1.3$, depth of field: $\sim 200$ nm).

Custom image analysis software allows us to track the $x$ and $y$ positions of any bead close to the objective focus plane. For a reliable analysis, special attention was brought to record only the beads far from the glass slide surfaces. 

For each tracer, the time-averaged mean-squared displacement (MSD) was calculated as follows: $<\Delta r^2(t)>= <(x(t+t')-x(t'))^2+(y(t+t')-y(t'))^2>_{t'} $. For a purely viscous fluid, the ensemble-averaged MSD increases linearly with the lag time $t$, according to: $<\Delta r^2(t)> = 4 D t$ (in two dimensions), where $D$ is the diffusion coefficient. In this case, the viscosity $\eta$ can be estimated with the Stokes–Einstein relation, $D=kT/6 \pi R \eta$, with $R$ the diameter of the bead and $kT$ the thermal energy \cite{Einstein1905}. 
Glycerol/water and ethylene glycol/glycerol solutions studied in section \ref{sec:FH}, were homogeneous and purely viscous, allowing for a simple determination of the sample viscosity. 

\subsection{Fluorescence microscopy and segmentation of red blood cells} \label{MM-Fluo}
\begin{figure*}[ht]
 \centering
 \includegraphics[width=16cm]{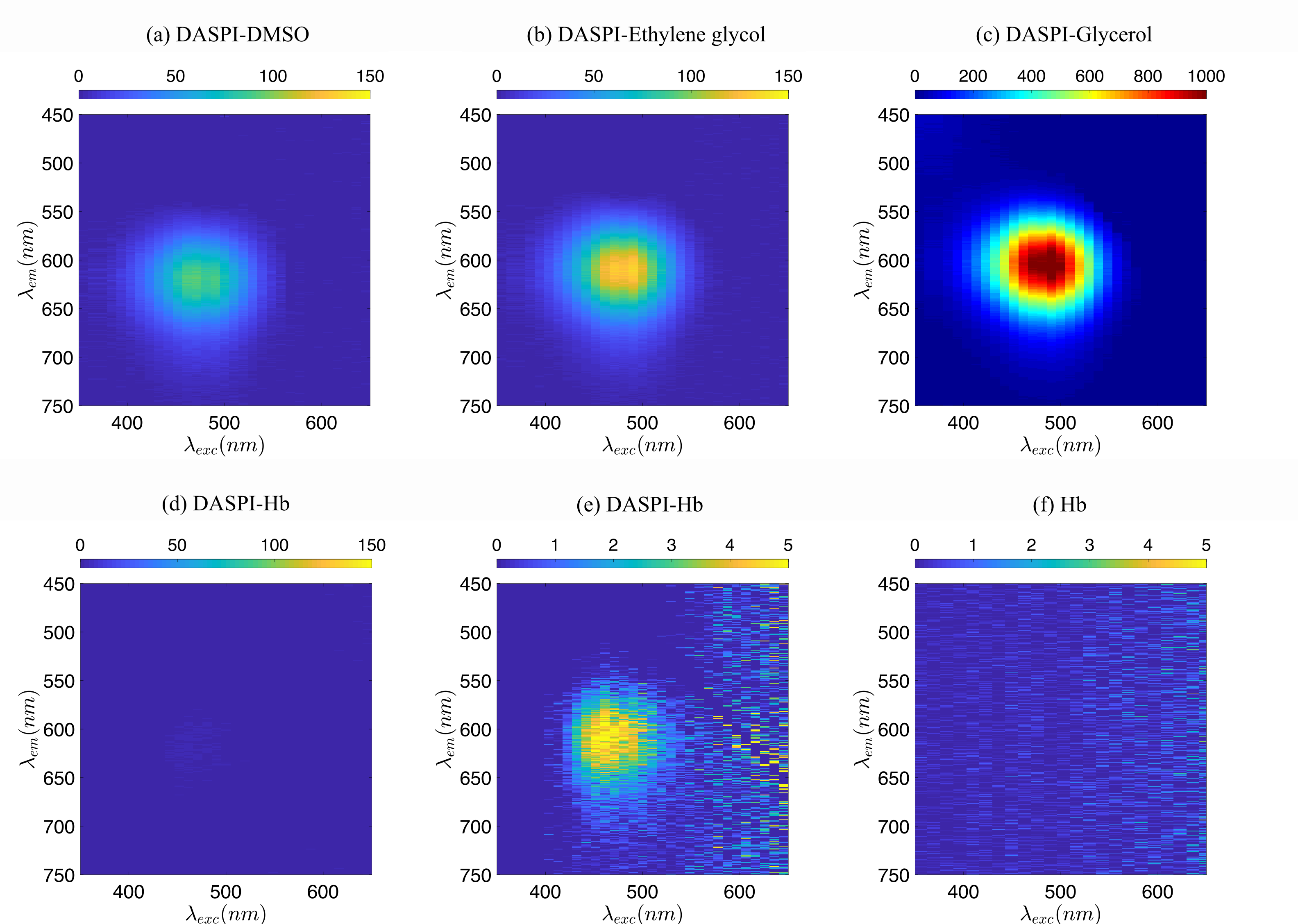}
\caption{Excitation-emission matrices of the DASPI molecular rotor at room temperature, at concentration $3.3$ $\mu$M in (a) DMSO ($\eta=2$ mPa.s), (b) Ethylene glycol ($\eta=13.5$ mPa.s), (c) Glycerol ($\eta=945$ mPa.s) and (d,e) Hemoglobin in PBS (1000x dilution: concentration $\sim 33$ mg/dL), different scale, $\eta=1$ mPa.s). The Fluorescence EEM of hemoglobin alone (f) was also measured to give evidence that there is no significant contribution of hemoglobin in the range of interest for DASPI. As can be seen, the fluorescence emission intensity (a,b,c) generally increases with the solution viscosity, measured in situ with microrheology experiments. In hemoglobin solutions however (d,e), the emission peak value is lower than in DMSO for similar viscosities because of the high absorbance of the solution and the behaviour of the rotor in an aqueous protein medium. 
}
\label{fig:matrix}
\end{figure*}

For fluorescence measurements of the DASPI molecular rotor in the solutions (water/glycerol, glycerol/ethylene glycol and hemoglobin) and RBC suspensions, samples were placed between a glass slide and a coverslip separated by a $250\,\mu$m thick spacer. For RBCs, the glass slides and coverslips were pre-treated with a plasma cleaner to prevent the formation of echinocytes.

Observation of the samples was carried out by fluorescence microscopy (Leica DMI 8, Leica Microsystems GmbH, Wetzlar, Germany) at $100X$ magnification with excitation between $460-500$ nm and detection between $565-625$ nm. The images were recorded with a fast sCMOS camera (OrcaFlash4.0 v2+, Hamamatsu Photonics France S.A.R.L., Massy, France). The intensity measurements of the fluorescence level  of the images were performed using ImageJ software (ImageJ, Rasband, W.S., ImageJ, U. S. National Institutes
of Health, Bethesda, Maryland, USA, https://imagej.nih.gov/ij/, 1997-2018). A macro program allows us to detect and segment RBCs on an image, measure their average fluorescence intensity, their surface area and their circularity. The circularity parameter is adjusted to select RBCs according to their orientation (front or side view) and to exclude echinocytes. 
In the experiments described in section \ref{sec:stiffening}, the segmentation is designed to measure the intensity on an area slightly smaller than the actual RBC surface area, in order to eliminate membranes contributions at the edges. During the analysis, the background fluorescence level is subtracted from the total image. This normalization allows us to obtain the signal actually produced by the red blood cells and to compare the realizations obtained when the temperature is varied.  

\subsection{Confocal microscopy} 

The penetration and localization of the DASPI molecular rotor in red blood cells was characterized with a Leica TCS SP8 confocal microscope equipped with a 40X oil-immersion objective ($1.3$ numerical aperture). 
Diluted red blood cell samples (hematocrit $0.5 \%$), incubated  with DASPI, were placed in a PDMS microchannel on the microscope stage. Confocal slices were acquired with a $0.5$ $\mu$m step in the $z$-direction to measure the fluorescence signal at different levels in RBCs. The excitation wavelength was $488$ nm and the depth of field was $600$ nm for a pinhole size of $63.5\,\mu$m. The laser power settings and detection gain were kept constant between experiments for comparison purposes. 
Image processing allowed us to extract fluorescence intensity profiles, averaged over a region of interest in the $(x,y)$ plane, for different $z$. 

\
\subsection{Spectrophotometry}
Fluorescence Excitation Emission Matrices (EEMs) of DASPI in various solvents (DMSO, Glycerol, Ethylene Glycol) and hemoglobin solutions were measured with a spectrophotometer at room temperature. The excitation wavelength was set between $350$ nm and $650$ nm and the emission spectrum was scanned between $450$ nm and $750$ nm. The excitation and emission slits were $5$ nm, the scan speed was $600$ nm/min and the voltage of the PMT detector was $600$ V. Absorption spectra were measured with a Varian Cary 50 UV-Vis spectrophotometer (Agilent) at room temperature. All experiments were performed in $2$ mL PS spectro cells (LP ITALIANA SPA). Hemoglobin was highly diluted to form a solution of reduced absorbance (1000x dilution $\sim 33$ mg/dL concentration). DASPI concentration was set to  $3.3\,\mu$M and the solutions were prepared with an $8$ mM DASPI stock solution.

\section{Results}

\subsection{DASPI molecular rotor properties}
\label{MRselection}
\subsubsection{Spectral properties}
\label{spectra}
\begin{figure}[ht]
\centering
  \includegraphics[width=8cm]{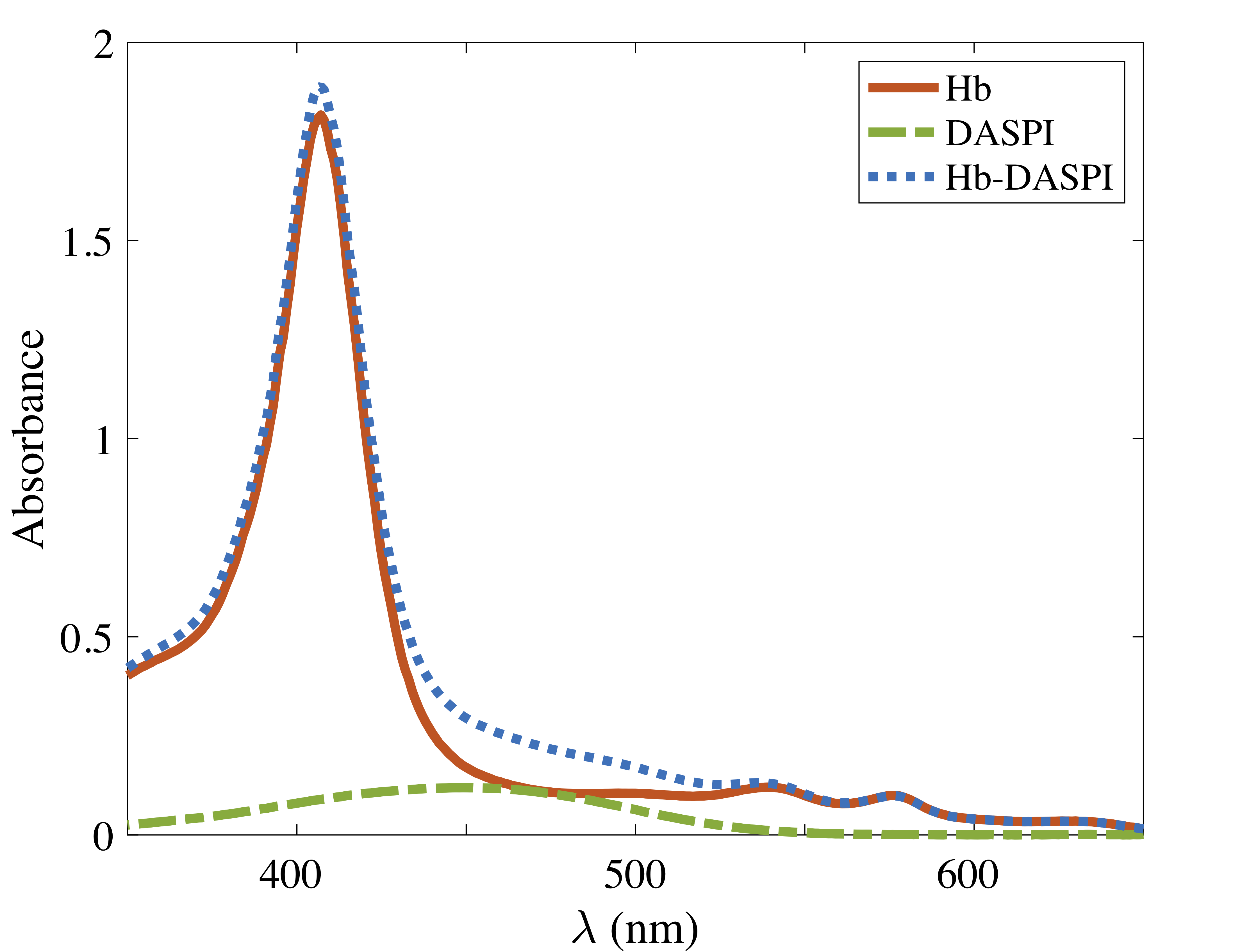}
  \caption{Absorbance spectra of the solutions (a, e, f) were also measured to (i) confirm the presence of compounds (Hb, DASPI) that were highly diluted to minimize hemoglobin absorbance, (ii) quantify the influence of compounds in the DASPI spectral range. }
\label{fig:abs}
\end{figure}

\begin{figure}[!ht]
\centering
\includegraphics[width=7cm]{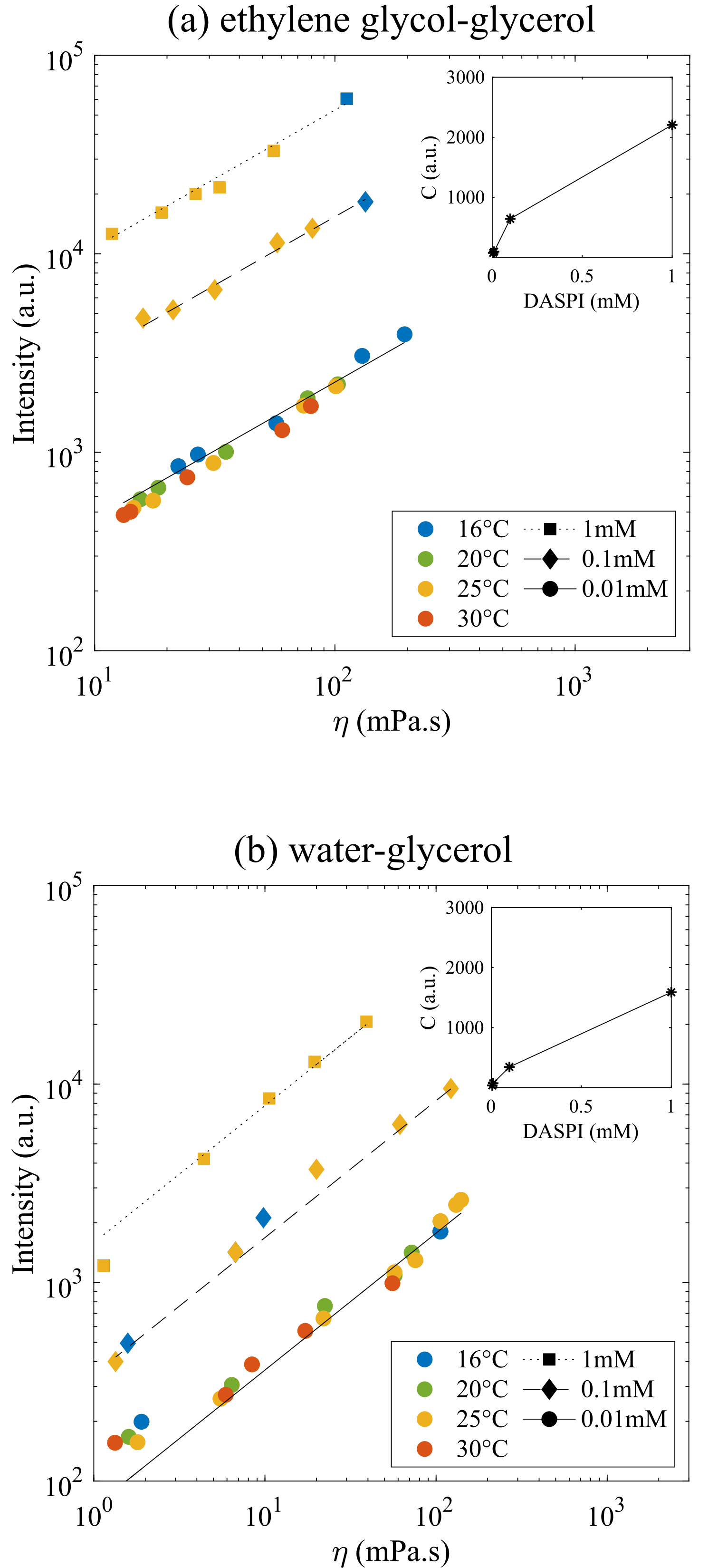}
\caption{Emission peak intensity (fluorescence microscopy, excitation at $488$ nm, emission $595\pm 30$ nm) of the molecular rotor DASPI at concentration $0.01$ mM, $0.1$ mM, $1$ mM in (left) ethylene glycol/glycerol solutions and (right) aqueous glycerol solutions with increasing viscosity. The solutions temperature was varied between $16$ and $20^{\circ}$C. The fluorescence intensity increases with the solution viscosity according to the Förster-Hoffmann equation $\log \phi= C + b \log \eta$, with solvent-dependent parameters $b$ and $C$ ; Ethylene glycol/glycerol solutions: $C= 94,641,2206 \pm $ and $b=0.69\pm 0.01$ and aqueous glycerol solutions: $C= 74,344,1590 \pm $ and $b=0.70\pm 0.01$.}
\label{fig:Forster}
\end{figure}

We selected the DASPI molecular rotor  (trans-4-[4-(dimethylamino)-styryl]-1-methylpyridinium iodide, 336408 from Aldrich) from the stilbene group\cite{Haidekker2010-revue} for its fluorescence characteristics. These are decoupled from those of hemoglobin known to have a broad absorption spectrum and intrinsic fluorescent properties in the ultraviolet\cite{Waterman1978}. 

Fluorescence Excitation Emission Matrices of DASPI were measured in four selected solvents, at room temperature: DMSO, Ethylene Glycol, Glycerol and hemoglobin/PBS solutions (dilution 1000x). 
Fig.~\ref{fig:matrix}-(a,b,c,e) shows that the fluorescence spectra of DASPI are similar from one solvent to another.
In the investigated range, DASPI is a single band emission molecular rotor, with fluorescence excitation and emission peaks respectively around $480$ nm and $600$ nm. More importantly, the excitation-emission spectrum of DASPI in hemoglobin (Fig.~\ref{fig:matrix}-(d,e)) is outside the intrinsic fluorescent spectrum of hemoglobin known to be in the ultraviolet with excitation maxima below $400$ nm and emission below $460$ nm\cite{Waterman1978}. 
The Fluorescence EEM of hemoglobin alone was also measured to demonstrate that there is no significant contribution of hemoglobin in the DASPI spectral range (spectrum (f)). 

From the EEM spectra (a,b,c) in Fig.~\ref{fig:matrix}, we can observe that the rotor has a fluorescence emission intensity which increases with viscosity. In hemoglobin solutions however (spectra (d,e)), the emission peak value is lower than in DMSO (spectrum (a)) for similar viscosities due to the high absorbance of hemoglobin and the behaviour of the rotor in an aqueous protein medium. This absorbance remains non-negligible even if the emission of DASPI is around a minimum of local absorption of hemoglobin as shown in Fig.~\ref{fig:abs} around $\sim 600$ nm. The absorbance spectra of Fig.~\ref{fig:abs} also attest to (i) the presence of compounds (Hb, DASPI) that have been highly diluted to minimize the absorbance of hemoglobin, (ii) the influence of these compounds in the spectral range of DASPI.

\subsubsection{Förster-Hoffmann relations for viscosity} \label{sec:FH}

A second step was to evaluate the sensitivity of DASPI to the local viscosity of the surrounding medium. In general, the photo physical sensitivity of fluorescent MRs depends on their interactions with the environment, which are influenced by viscosity, polarity and solubility\cite{haidekker2005,Haidekker2010-revue,howell2011}. Temperature is also a parameter that can affect the rate of TICT formation, and thus influence the parameter $C$ of the Förster-Hoffmann equation \cite{law1980,kung1986,iwaki1993,haidekker2000}.

The emission intensity of DASPI was measured by fluorescence microscopy (excitation at $480 \pm20$ nm, emission at $595\pm 30$ nm) in ethylene glycol/glycerol solutions and aqueous glycerol solutions whose viscosity was measured in situ by microrheology, just before the fluorescence intensity measurements.
We measured the emission intensity of DASPI at different temperatures between $16$ and $20^{\circ}$C and at different rotor concentrations: $0.01$, $0.1$ and $1$ mM (Fig.~\ref{fig:Forster}-(a,b)). The relative concentration of the two solvents in each solution was modified to extend the achievable viscosity range, typically between $2-200$ mPa.s.
Ethylene glycol/glycerol solutions were chosen as one of the study systems because their polarity remains approximately constant when the relative concentration between the solvents and therefore the viscosity varies, as the polarities of the two liquids -- glycerol and ethylene glycol -- are similar\cite{haidekker2005}.
In aqueous solutions of glycerol however, the polarity of the solvent decreases with increasing viscosity, based on the dielectric constant values of water and glycerol (water: $80.1$ F/m, glycerol: $42.5$ F/m, ethylene glycol: $37.7$ F/m)\cite{haidekker2005}.

\begin{figure*}[ht]
\centering
\includegraphics[width=16cm]{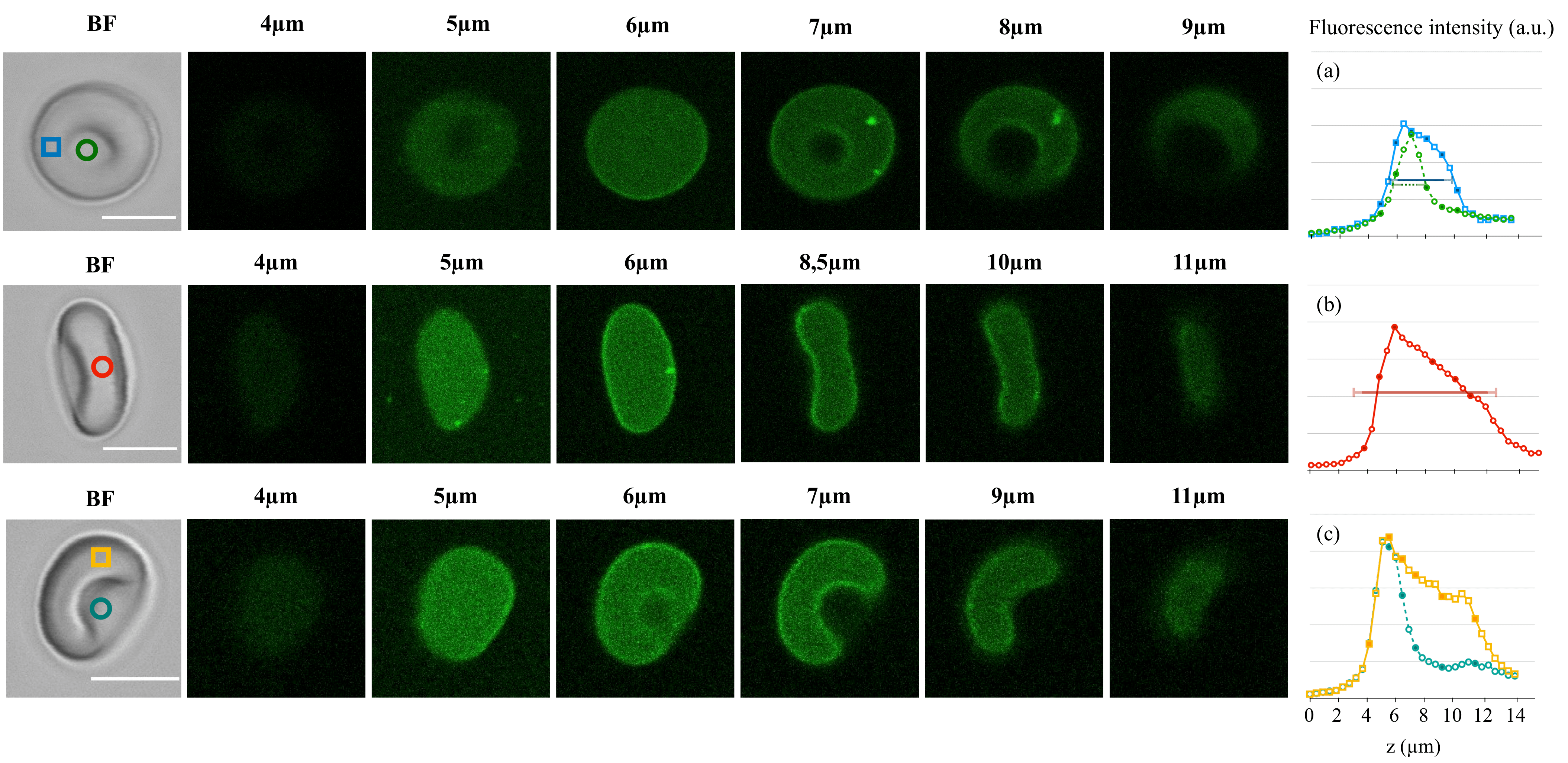}
\caption{Bright field image and confocal $z$-stacks (confocal microscopy, X40 objective, excitation at $488$ nm, detection between $542-626$ nm, scale bar: $5\,\mu$m) for three healthy RBCs with different orientations in a $1$ mM DASPI solution. The $z$ position of the slices is indicated. On the right, the fluorescence intensity profiles in the $z$ direction are displayed. The mean intensity was measured in the depicted color boxes, yielding the fluorescence intensity profiles in the $z$-direction displayed on the right. Each data point corresponds to a confocal slice ($0.5\,\mu$m distance). The horizontal lines on the intensity profiles correspond to the typical sizes expected for a RBC with 0.6 $\mu$m (depth of field) added on each side. (a) Expected thickness at the edges: $3 + 2*0.6$ $\mu$m (blue profile); in the biconcave zone: $1 + 2*0.6$ $\mu$m (green profile). (b) Expected diameter: $8 + 2*0.6$ $\mu$m (red profile).}
\label{fig:confocslices}
\end{figure*}

Fig.~\ref{fig:Forster} shows that the rotor's emission intensity increases with the local viscosity and follows a Förster-Hoffmann equation, $\log \phi= C + b \log \eta$ in both solutions, with $\phi$ the quantum yield directly proportional to the emission intensity,
%évident sinon ref Haidekker2010-revue\cite{Haidekker2010-revue}}
and $b$ and $C$ two solvent-dependent parameters.
The  $C$ parameter increases significantly with the DASPI concentration as shown in the insets in the Fig.~\ref{fig:Forster}. The exponent $b$ was similar in the two solutions and close to $0.7$.
(water/glycerol: $C= 74,344,1590$ and $b=0.70\pm 0.01$; ethylene glycol/glycerol solutions: $C=94,642,2206 $ and $b=0.69\pm0.01 $). 
From our measurements, we deduce that temperature has little influence on 
the $C$ parameter, and thus on the signal emitted by  DASPI, as highlighted specifically at the $0.01$ mM rotor concentration where all data points coming from different temperatures are on the same Förster-Hoffmann power-law curve, in consistency with a previous study \cite{howell2011}. 
Similarly, polarity has little effect on rotor response when comparing ethylene glycol/glycerol solutions of constant polarity and aqueous glycerol solutions of variable polarity, which is also consistent with previous studies where molecular rotors with a single emission band showed an emission intensity strongly dependent on viscosity but not on solvent polarity\cite{kung1986,haidekker2005,allen2005}. At low viscosity around $1-2$ mPa.s in the aqueous glycerol solutions, a deviation is observed from the Förster-Hoffmann law as already observed in previous studies \cite{Haidekker2010-revue,haidekker2005}.

\subsection{Molecular rotor DASPI in red blood cells}

\subsubsection{Penetration in healthy red blood cells \label{sec:confoc}}

The penetration of the DASPI molecular rotor into healthy red blood cells was studied by confocal microscopy. Images of cells with different orientations were taken during their slow sedimentation in the observation microchannel. 
Fig.~\ref{fig:confocslices} shows stacks of slices for three different RBCs, corresponding to different focus planes in the $z$-direction. We observed that the membrane fluorescence signal is slightly stronger, certainly due to the interaction between the molecular rotor and the membrane components. 
Inside the cell (cytosol), DASPI fluorescence is homogeneous in the observation plane, regardless of the $z$-position and orientation of the cell. 
Confocal microscopy allows us to unambiguously separate the fluorescence response from the membrane and cytosol. Indeed, as soon as the upper and lower membranes are out of the depth of field ($\sim\,600$ nm), they no longer contribute to the field of view. This indicates that the fluorescence signal observed inside the cell, far enough from the membrane, comes only from the cytosol, without any signal from the membrane. 
The molecular rotor is thus uniformly distributed throughout the volume of the red blood cells.
Further evidence of the uniform and effective penetration of the molecular rotors into  red blood cells is provided by the intensity profiles of these red blood cells in the $z$ direction. By scanning  different parts of the red blood cells, we show that the DASPI fluorescence signal is present throughout the entire red blood cell, over a distance that corresponds approximately (given the depth of field) to the known typical sizes of the red blood cell: diameter $8\,\mu$m (Fig.~\ref{fig:confocslices}b), thickness at the edge $3\,\mu$m (Fig.~\ref{fig:confocslices}a), thickness in the biconcave zone $\sim 1\,\mu$m (Fig.~\ref{fig:confocslices}a). 

\subsubsection{Sensitivity of DASPI fluorescence to red blood cell stiffening} \label{sec:stiffening}

\begin{figure}[ht]
\centering
\includegraphics[width=8cm]{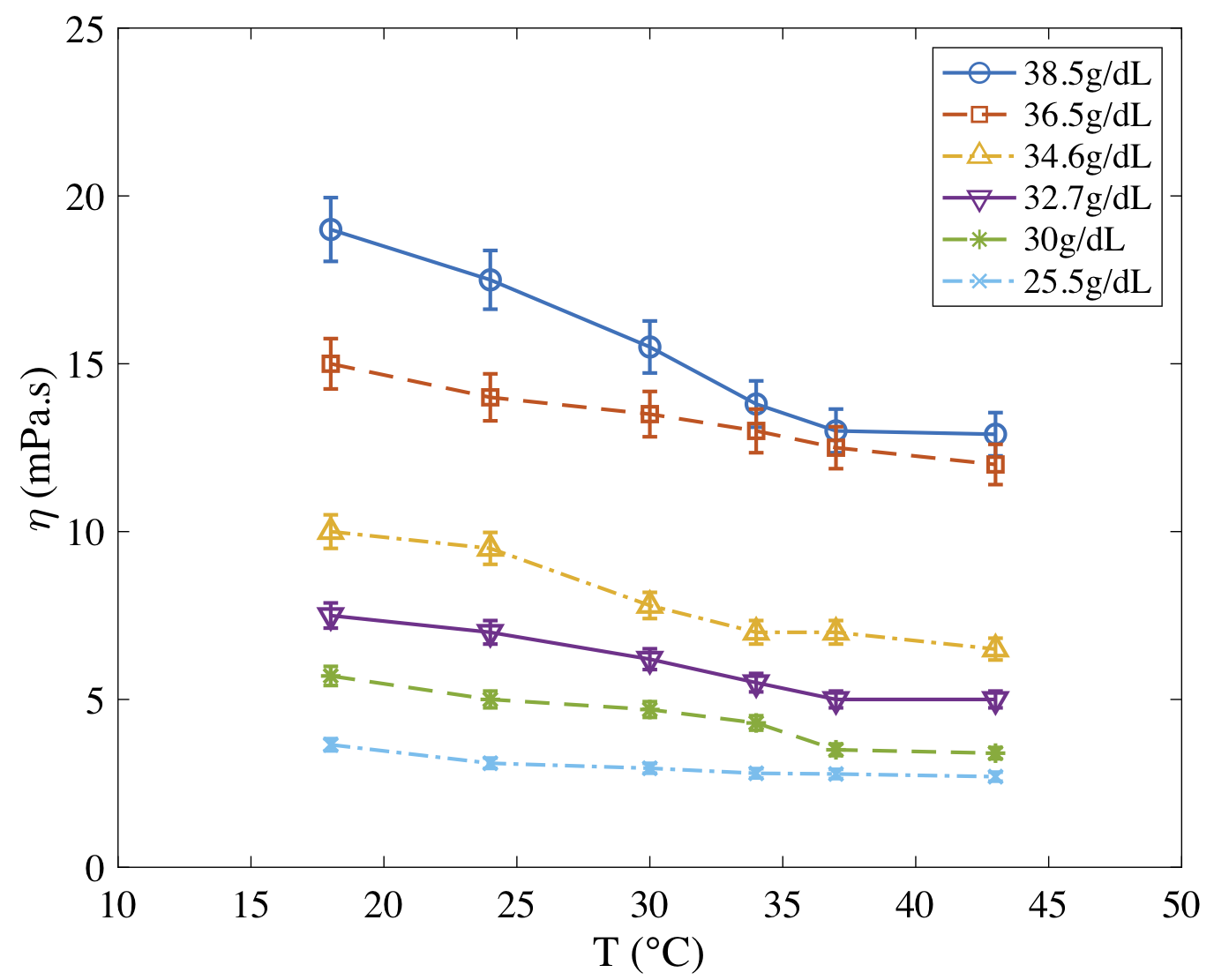}
\caption{Viscosity of hemoglobin solutions as a function of temperature, measured by microrheology. Hemoglobin concentration ranges from $25.5$ g/dL to $38.5$ g/dL. Viscosity decreases with temperature and increases with concentration. For a given concentration range, viscosity values are spread over a wider range at low temperature than at high temperature. }
\label{fig:Hb-T}
\end{figure}

\begin{figure*}[ht]
\centering
\includegraphics[width=17cm]{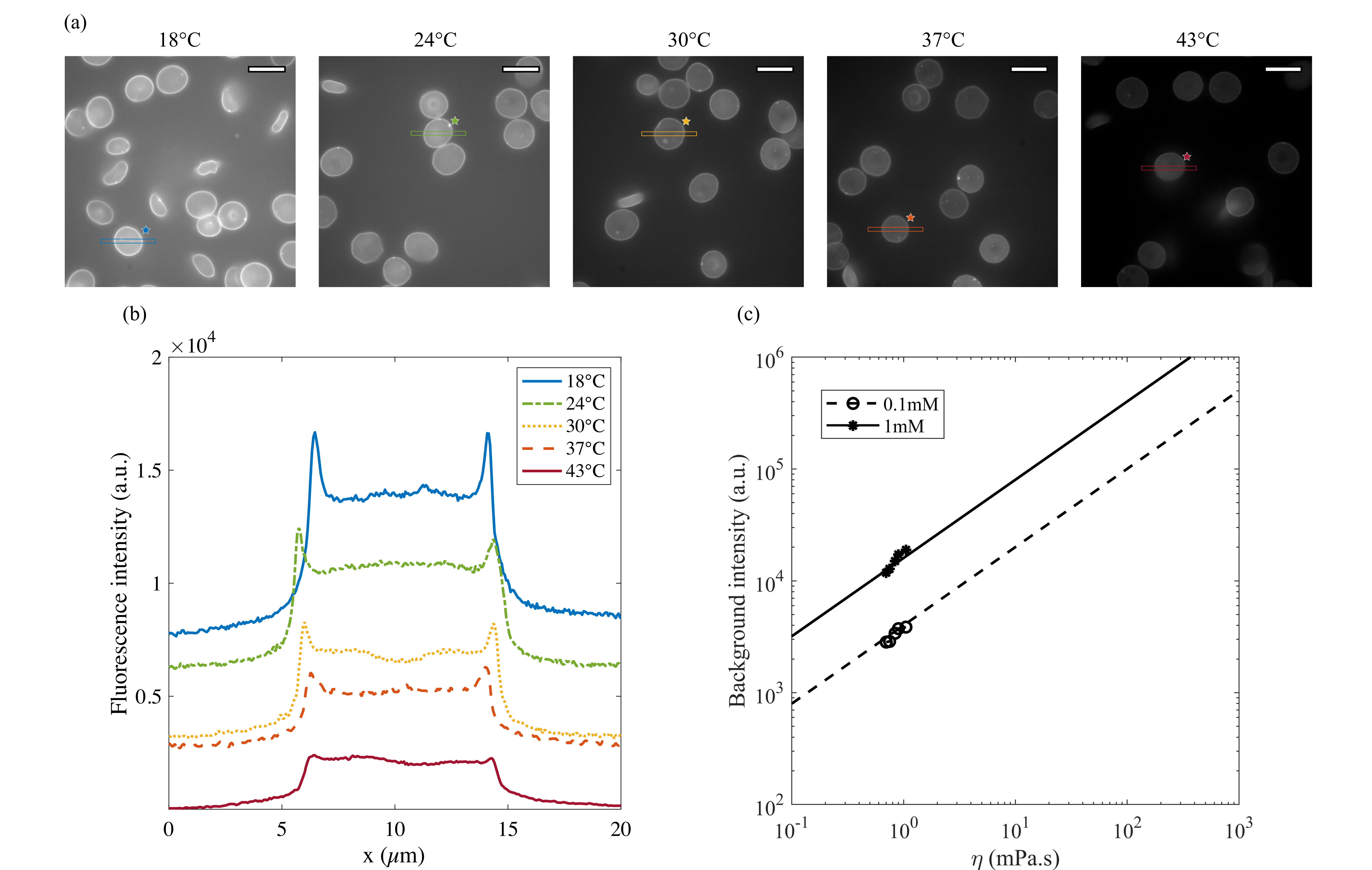}
\caption{Fluorescence microscopy of red blood cells 
at different temperatures (excitation $480\pm20$ nm, emission $595\pm30$ nm). (a) Images of RBCs with increasing temperature in a $1$ mM DASPI solution (scale bar: $10\,\mu$m); (b) fluorescence intensity profiles of the  RBCs labelled with a star along the x-axis. Fluorescence was averaged over a rectangle of width $1.5\,\mu$m. (c) Background fluorescence intensity with viscosity in DASPI solutions at $0.1$ mM and $1$ mM. The data were adjusted with a Förster-Hoffman equation with exponent $x=0.7$.}
\label{fig:GR-T}
\end{figure*}

\begin{figure}[!ht]
\centering
\includegraphics[width=8cm]{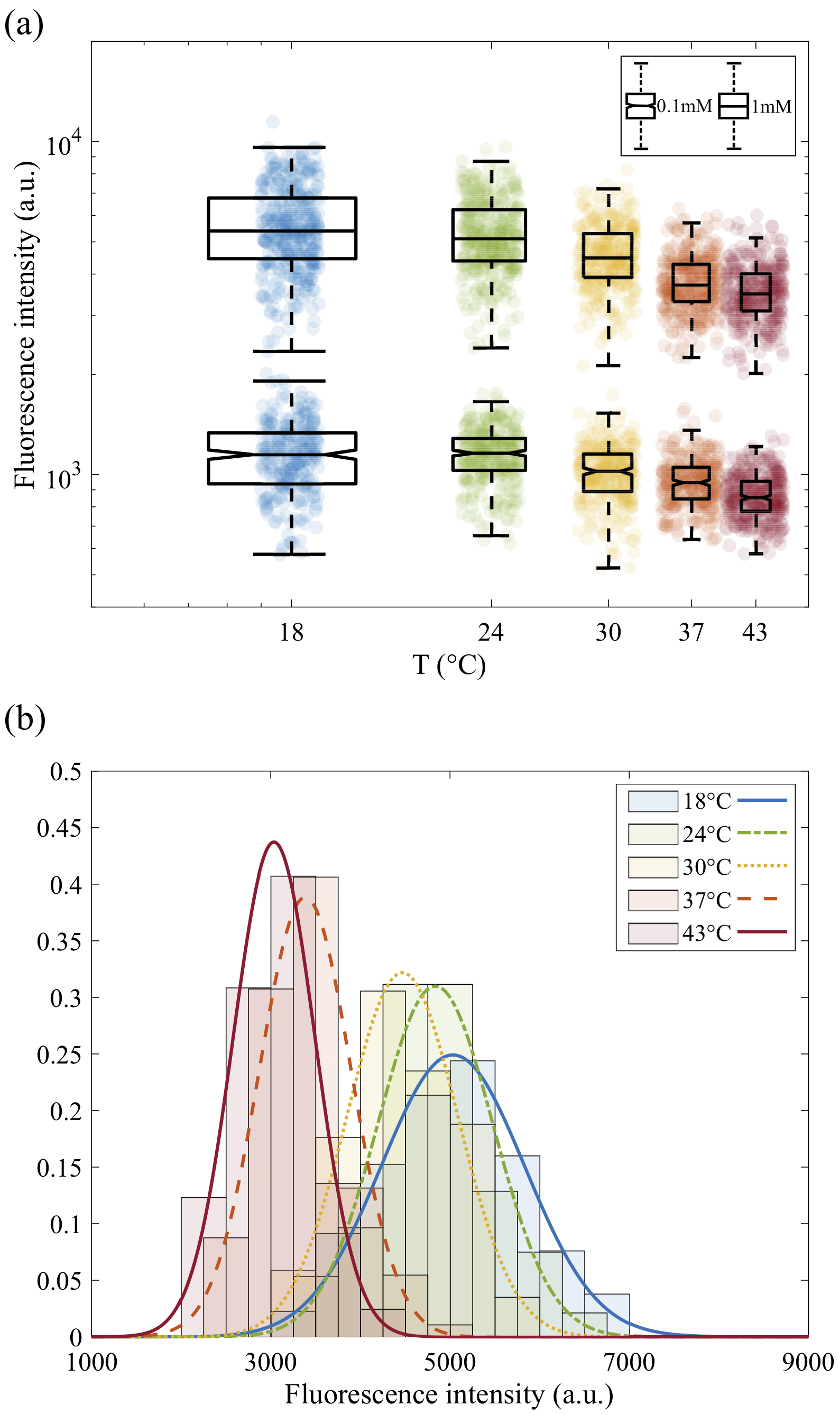}
\caption{(a) Fluorescence intensity of RBCs as a function of temperature in $0.1$ mM and $1$ mM DASPI solutions. RBCs come from five donors with MCHCs between $30$ g/dL and $35$ g/dL. Approximately four hundred cells were imaged at each temperature, and their average intensity was measured; (b) Normalized histogram plots and density functions of RBCs fluorescence intensity in 1 mM DASPI solution at different temperature, 
for a single donor with a MCHC of $33.5$ g/dL. The histogram plots were calculated from approximately $80$ data points.}
\label{fig:GR-T-hist}
\end{figure}

Lowering the temperature is known to increase the rigidity of red blood cells at two levels: the viscosity of hemoglobin, main component of the cytosol, increases \cite{artmann1998,kelemen2001} and the membranes become stiffer\cite{lecklin1996,xu2018,waugh1979}. In this work, we use temperature as a parameter to control the stiffness of RBCs. It was varied between $18^\circ$C and $43^\circ$C.
 
Independent measurements of the viscosity of hemoglobin solutions as a function of temperature are shown in Fig.~\ref{fig:Hb-T}. Viscosity was measured by microrheology in hemoglobin solutions with concentrations between $25.5$ g/dL and $38.5$ g/dL. Our experiments show that hemoglobin behaves like a purely viscous fluid (not shown) and that the solutions are homogeneous (not shown) and that the viscosity decreases with temperature at fixed concentration, and increases with concentration at fixed temperature. For a given concentration range, viscosity values extend over a wider range at low temperature than at high temperature. This evolution of viscosity with temperature is consistent with the literature\cite{artmann1998,kelemen2001}.

Fig.~\ref{fig:GR-T} shows images of red blood cells at different temperatures (Fig.~\ref{fig:GR-T}-a) as well as the intensity profiles of the cells labelled with a star (Fig.~\ref{fig:GR-T}-b). The intensity profiles all show homogeneous backgrounds, sharp peaks at the edges and a plateau inside the cell. Both the background (i.e. PBS solution) and the RBC profiles increase as the temperature decreases. Since  DASPI signal does not depend on temperature in the studied range (section \ref{sec:FH}), this increase in the peak and plateau fluorescence when temperature decreases demonstrates that the fluorescence signal of DASPI is sensitive to, and increases with, the rigidity of RBCs.
The observed peaks on the profiles are further evidence, along with confocal microscopy observations, that the membrane-DASPI interactions result in a stronger fluorescence response than that of cytosol-DASPI. A closer look at the plateaus (for example clearly visible here at $30^{\circ}$C) reveals small height variations that reflect the cell geometry and suggest that the cytosol contribution is proportional to the local thickness of the cell. 
The viscosity-intensity dependence of the background was determined at each temperature using microrheology in the PBS surrounding the RBCs. Fig.~\ref{fig:GR-T}-c shows that the rotor emission increases with the PBS viscosity and that the data can be adjusted with a Förster-Hoffman power law of exponent $x=0.7$\footnote[1]{Data can be fitted with a Förster-Hoffman relation even at low viscosity were a deviation from the law at higher viscosity was observed (section \ref{sec:FH}).}.

To ensure a statistically relevant sampling, we studied blood samples from five different donors with mean corpuscular hemoglobin concentrations (MCHC) between $30$ g/dL and $35$ g/dL. About four hundred cells were imaged at each temperature, and the average intensity was measured on each cell (see section \ref{MM-Fluo}). Fig.~\ref{fig:GR-T-hist}-a gathers the data together within a boxplot. As the temperature decreases, the median fluorescence intensity increases (by $35\%$ in $1$ mM DASPI solutions and $26\%$ in $0.1$ mM DASPI solutions) and the collected data are more scattered. The same pattern is observed when considering RBCs coming from a single donor (Fig.~\ref{fig:GR-T-hist}-b): the fluorescence intensity distribution of RBCs shifts towards high intensities and becomes wider at low temperatures. 
This dispersion of red blood cell fluorescence intensities, measured at a given temperature, reflects the variability of their properties: variation in the mean corpuscular hemoglobin concentration from one donor to another (Fig.~\ref{fig:GR-T-hist}-a) and dispersion due to the age of the cells within a sample (Fig.~\ref{fig:GR-T-hist}-a-b). Variations in the physico-chemical properties of membranes may also play a role in the dispersion of the data, the separation of the two contributions will be discussed in the following section. However, these experiments prove that the DASPI molecular rotor can quantify the rigidity of RBCs as well as the dispersion, or heterogeneity, of the mechanical properties of red blood cells in a sample.

 \begin{figure*}[!ht]
\centering
\includegraphics[width=15cm]{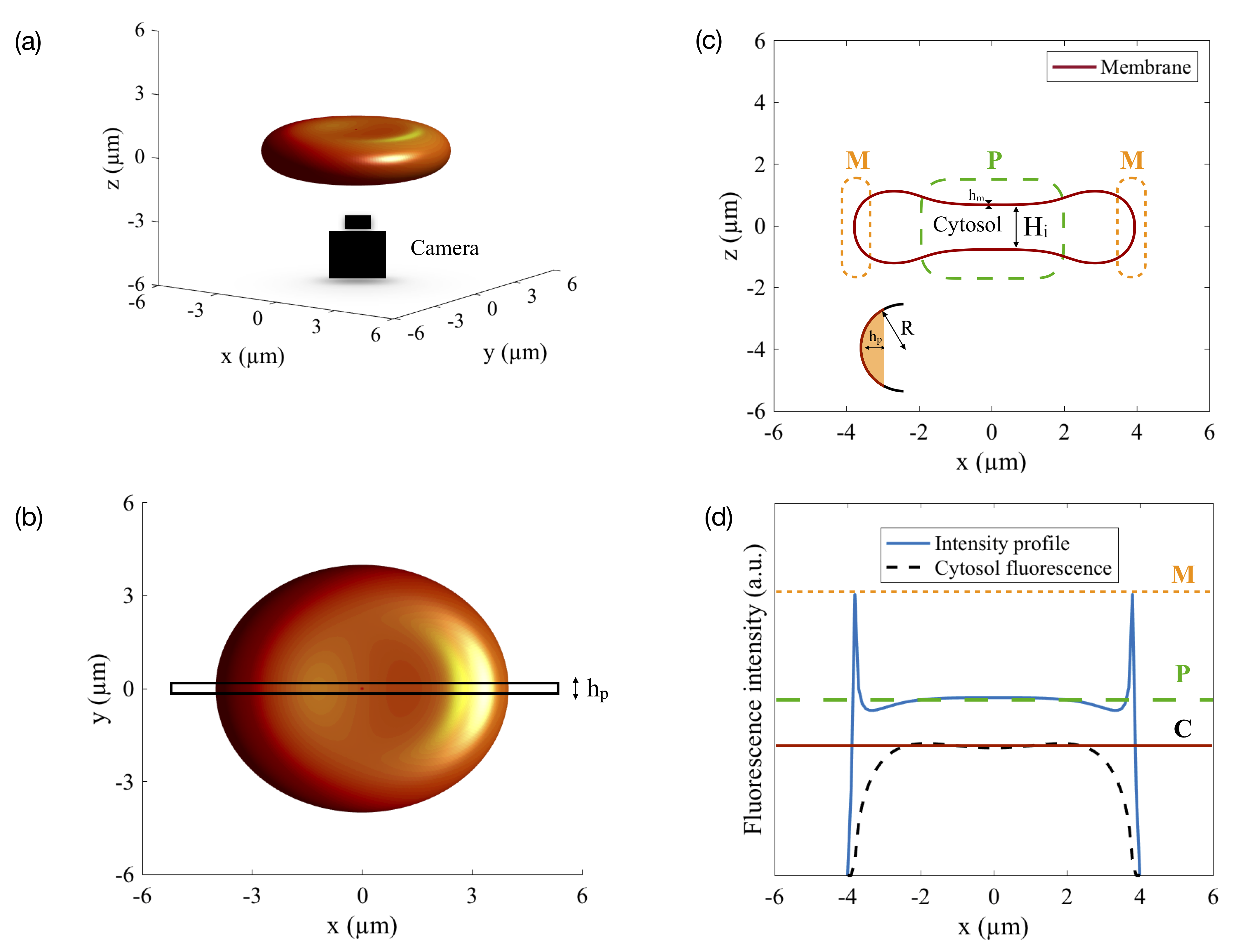}
\caption{Model RBC and characteristic intensity profile obtained in fluorescence microscopy. (a) 3D representation of the RBC; (b) 2D representation of the RBC  and measurement area of the intensity profile (rectangle of width $h_p$ centered around $y=0$); (c) 2D section of the model RBC for $y=0$, the membrane is of constant thickness $h_m$ and the cytosol is of constant height at its center $H_i$; zooming at the edges of the RBC, we define the average radius of curvature $R$, and the membrane (in red) and cytosol (in orange) portions contributing to the peak signal $M$ over the width $h_p$; (d) corresponding intensity profile and possible trend for the cytosol fluorescence, the intensities of the characteristic peaks ($M$) and of the plateau ($P$) can be measured on the profile, they correspond respectively to the edges of the RBC (zoomed above) and to the center part which in a simplified model is of constant height. The cytosol fluorescence $C$ in the central part is also shown in the figure.}
\label{fig:prof-annexe}
\end{figure*}
\subsection{Estimation of cytosol fluorescence 
%with a simplified RBC model
using a simple model of RBC}

\begin{figure*}[ht]
\centering
\includegraphics[width=17cm]{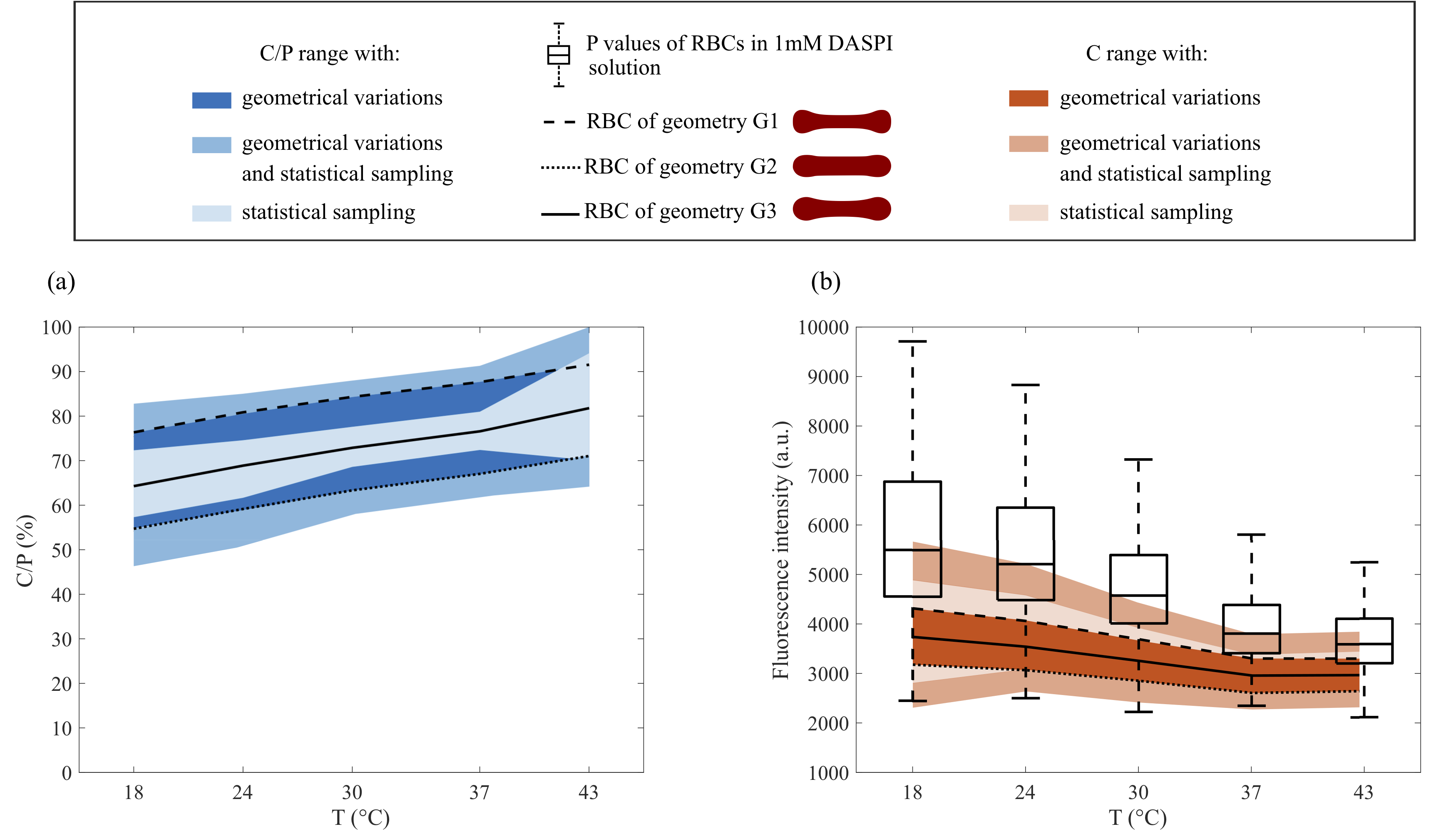}
\caption{Cytosol to plateau $C/P$ (a) and cytosol $C$ (b) contributions with temperature, for RBCs in a $1$ mM DASPI solution. The fluorescence intensities of RBCs with temperature from Fig.~\ref{fig:GR-T-hist}-a (boxplot, corresponding to $P$) are reported in (b). The light-colored zones delineate an area corresponding to the statistical sampling effect. The continuous lines within these light color zones are obtained by considering the median values calculated with the G3 geometry ($R=1.25\,\mu$m, $H_i= 1.6\,\mu$m). The dark bands quantify the dispersion of $C/P$ and $C$ for all possible geometries in the ranges $1\,\mu$m $<R< 1.5\,\mu$m and $1.3\,\mu$m $<H_i< 1.9\,\mu$m. The dashed lines correspond to the median values of $C$ and $C/P$ obtained for the extreme geometries G1 and G2. The intermediate color bands gather both the effects of statistical sampling and the effects of possible geometries.}
\label{fig:geom-annexe}
\end{figure*}

In fluorescence microscopy experiments (section \ref{sec:stiffening}), both the membrane and the cytosol contribute to the signal measured on RBCs. We propose here 
a simplified RBC model to separate the two contributions and to find the evolution of the cytosol fluorescence with temperature (i.e. stiffness). 
%x
We show how geometrical parameters can influence this estimation  and that we can in principle relate fluorescence measurements to the mechanical properties of the cytosol, i.e. the internal viscosity of hemoglobin.

\subsubsection{Simplified RBC model}
In section \ref{sec:stiffening}, all imaged RBCs were chosen with the same front view orientation.
%In a simplified model, we consider RBCs of constant height in their center 
In a simplified model, we consider RBCs with constant height in their center and curved edges, as represented in Fig.~\ref{fig:prof-annexe}-(a,b,c). The camera being along the $z$ axis, the fluorescence image obtained is a projection in the $(x,y)$ plane, 
%squared into pixels of surface $h_p^2$
%marked by 
consisting in a matrix of squared units of area $h_p^2$. In order to take into account the effect of diffraction, $h_p$ is defined as the optical resolution of the microscope, i.e. the radius of the Airy disk: $h_p=1.22\cdot\lambda/\left(2\cdot NA\right)=277$ nm. We focus on the fluorescence intensity profile obtained along the $x$ axis, averaged over the rectangle centered in $y=0$ and height $h_p$ (Fig.~\ref{fig:prof-annexe}-(b,d)).
A typical intensity profile is shown in Fig.~\ref{fig:prof-annexe}-d with two peaks at the edges and a plateau at the center. Assuming that the absorbance of hemoglobin has little effect on the considered lengthscales, the intensity represented in Fig.~\ref{fig:prof-annexe}-d is the integral of the intensities over $z$. 
In Fig.~\ref{fig:prof-annexe}-c, the corresponding 2D section of the RBC in the $(x,z)$ plane is illustrated.
The two peaks of intensity $M$ correspond to the curved edges, where the membrane dominates and projects a higher fluorescence intensity; the plateau of intensity $P$ corresponds to the central part of the cell, which is, in this model, a cytosol of constant height with two membranes, top and bottom. Peak and plateau intensities ($M, P$) can both be measured on the intensity profiles (Fig. \ref{fig:prof-annexe}-d).
The cytosol fluorescence cannot be measured independently. 

However we show in Fig.~\ref{fig:prof-annexe}-d

that its contribution should be proportional to the local thickness of the cell. 
In the following, we estimate the fluorescence of the cytosol in the central part ($C$) and its contribution to the measurable plateau intensity ($C/P$).

The relevant parameters are as follows:

\begin{itemize}
\item{geometrical parameters:}
\begin{itemize}
    %\item $h_p$: pixel size;
    \item $h_m$: thickness of the membrane; 
    \item $H_i$: thickness of the cytosol along the $z$ axis in the central part; 
    \item $R$: average radius of curvature at the edges of the RBC;
\end{itemize}
\item{Fluorescence intensities:}
\begin{itemize}
     
    \item $\phi_m \left(\text{m}^{-3}\right)$: fluorescence intensity per unit volume of the membrane;
    \item $\phi_i\left(\text{m}^{-3}\right)$: fluorescence intensity per unit volume of the cytosol.
\end{itemize}
\end{itemize}
Cytosol and plateau intensities ($C$ and $P$) are related to the terms defined above as follows: 
\begin{equation}
\begin{aligned}
\label{eq1}
C & = h_p^2H_i \phi_i\\
P & = C+2h_p^2h_m\phi_m = h_p^2\left(H_i\phi_i+2h_m\phi_m\right)\\ 
\end{aligned}
\end{equation}

As $h_p$ is large compared to the membrane thickness $40$ nm$ \leq h_m \leq 100$ nm\cite{heinrich_elastic_2001,smith_myosin_2018,lux_anatomy_2016}, the peak signal $M$ is composed of the signal of the portions of the curved membrane and cytosol over the width $h_p$ (zoom in Fig.~\ref{fig:prof-annexe}-c), so that:
\begin{equation*}
M=h_mh_pS\Phi_m+h_pA\Phi_i\\ 
\end{equation*}
with
\begin{itemize}
    \item $S \simeq 2\sqrt{2Rh_p}$: curvilinear length of the membrane, in red on Fig.~\ref{fig:prof-annexe}-c-zoom;
    \item $A \simeq \frac{4h_p}{3}\sqrt{2Rh_p}$: area of cytosol, in orange on Fig.~\ref{fig:prof-annexe}-c-zoom.

\end{itemize}
This leads to the following formula for the peaks intensity: 
\begin{equation}
M = h_p^2\sqrt{\frac{2R}{h_p}}\left(2h_m\phi_m+\frac{4}{3}h_p\phi_i\right)\\ 
\end{equation}
Finally, the cytosol intensity reads: 
\begin{equation}
 %\frac{C}{P}=1-\frac{2}{n}\times\frac{M}{P}.
C = P\cdot \frac{1-\sqrt{\frac{h_p}{2R}}\frac{M}{P}}{1-\frac{4}{3}\frac{h_p}{H_i}}
 \label{eq2}
\end{equation}

In practical cases, $ h_p \ll H_i$ so that the denominator of equation \ref{eq2} cannot be zero. Equation \ref{eq2} indicates that $C$ depends on the geometrical parameters $R$ and $H_i$ and the values $M$ and $P$ which can be measured on the intensity profiles (Fig.~\ref{fig:prof-annexe}). $C$ can also be expressed, using equation \ref{eq1},
with the ratio of elementary intensities $\Phi_m/\Phi_i$: 
\begin{equation}
C = P\cdot \frac{1}{1+\frac{2h_m}{H_i}\frac{\Phi_m}{\Phi_i}}
 \label{eq4}
\end{equation}
In the following, we will focus on $C$ and $C/P$ which represents the contribution of cytosol to the plateau intensity $P$, measured in fluorescence microscopy experiments (section \ref{sec:stiffening}).
 
\subsubsection{Estimation of the cytosol fluorescence}
We estimate $C$ and $C/P$ with temperature, by first defining acceptable ranges for the geometrical parameters $R$ and $H_i$, then using equation \ref{eq2} for a statistical sampling of about $100$ RBCs, representative of the data dispersion in Fig.~\ref{fig:GR-T-hist}-a.
We also focus on unravelling the dispersion of data related to the intrinsic properties of RBCs (internal viscosity, membrane stiffness) and to the geometrical unknown ($R$ and $H_i$).

A RBC is classically described in the literature as biconcave, with a maximum height at the edges of 2-3$\,\mu$m, a radius of curvature of about half this maximum height, and a minimum height at the center of 0.8-1 $\mu$m\cite{evans1972}. Therefore, in our model, $R$ is chosen in the range $1\,\mu$m $\leq R \leq 1.5\,\mu$m and $H_i$ in the range $1.3\,\mu$m $\leq H_i\leq 1.9\,\mu$m as an average between the maximum and minimum heights. By varying $R$ and $H_i$ within these ranges, we cover a large number of realistic shapes for a RBC. We note G1 ($R=1.5\,\mu$m, $H_i=1.3\,\mu$m) and G2 ($R=1\,\mu$m, $H_i=1.9\,\mu$m) the geometries leading to extreme values of $C/P$ for a given $M/P$ (see equation \ref{eq2}). G1 corresponds to a thin RBC with sharp edges, and G2 to a thick RBC with rounded edges. We also note G3 ($R=1.25\,\mu$m, $H_i=1.6\,\mu$m) the geometry corresponding to the mean value of the range considered. G1, G2 and G3 are shown in Fig.~\ref{fig:geom-annexe}. In this section, we consider that for a given RBC, $R$ and $H_i$ do not vary with temperature. This hypothesis is supported by recent observations by Jaferzadeh et al.\cite{jaferzadeh2019} showing that the shape of the edges does not vary significantly with temperature, and that although fluctuations in the central part are more important at higher temperatures, the mean thickness does not vary.

We evaluated the $C/P$ and $C$ values, separating or not the effects of RBC geometry and statistical sampling. As indicated above, approximately $100$ RBCs per temperature were considered and the corresponding pairs ($M$,$P$) were measured on each of their intensity profiles. 
In Fig.~\ref{fig:geom-annexe}-b, the $P$ values from section \ref{sec:stiffening} were added for comparison. 
The first way of processing the data was to visualise the effect of the statistical sampling alone, when the  geometry is set at its mean value G3. It delineates light-colored zones corresponding to the data dispersion of $C/P$ and $C$ (Fig. \ref{fig:geom-annexe}-(a,b)). The continuous lines within these zones correspond to their median values. 
A second way of processing data was to quantify the effect of the possible geometries only.
To do this, the geometry was set at the G1 or G2 extremes. Next, the resulting median values of $C/P$ and $C$ were calculated and are represented by dashed lines. These dashed lines define dark bands that quantify the dispersion due to possible geometries.
Finally, both effects, statistical sampling and geometry, were combined, and are shown in the intermediate color bands (all possible geometries for $100$ RBCs).

In all cases, the bands corresponding to $C/P$ values increase roughly linearly with temperature. 
%Since $C/P$ increases with temperature, it can be noticed that the observed decrease of $C$ in Fig.~\ref{fig:geom-annexe}-b, is therefore related to the sharp decrease of $P$ values with temperature. 
In parallel, the ones corresponding to $C$ decrease with temperature, as evidenced in particular in Fig.~\ref{fig:geom-annexe}-b, when the geometry is fixed at the extremes or its mean value (dashed and solid lines). 
This decrease in the cytosol signal with temperature, caused by the sharp decrease of $P$, is consistent with the decrease of the viscosity of hemoglobin solutions measured in Fig.~\ref{fig:Hb-T}. 

We focus now specifically on the dispersion of $C$ and $C/P$ with temperature. 
The dispersion of $C/P$ remains constant when considering the combined effects of geometry and statistical sampling (intermediate color band). 
Conversely, the one of $C$ decreases with temperature. Since its light band is larger than the dark band which is quasi constant in width, this decrease is mainly due to the effect of statistical sampling alone.
It thus reflects the intrinsic heterogeneity of a blood sample and in particular the dispersion of internal viscosity. This is quite consistent with the decrease in the dispersion of hemoglobin viscosity values (independent measurements in section \ref{sec:stiffening}).

While more information on the cell geometry would be needed to reduce uncertainties in the determination of $C$, this simple model shows that it is in principle possible to separate the contributions of the cytosol and membrane from simple fluorescence profile measurements, and that the derived values are consistent with independent viscosity measurements on hemoglobin solutions.
The same study could be devoted to the membrane by considering $(P-C)/2 = h_p^2h_m\Phi_m$ which represents the intensity of a membrane element of surface area $h_p^2$ and thickness $h_m$ (equation \ref{eq1})\footnote[2]{We observe that the signal from the membrane and its dispersion due to statistical sampling increases with decreasing temperature (not shown).}. We can however note that the increase in $C/P$ reflects a decrease in the $\Phi_m/\Phi_i$ ratio (equation \ref{eq4}), which means that the contribution of the membrane decreases much faster than that of the cytosol when temperature increases, a feature that remains to be explored in terms of rotor-membrane interactions.

\section{Discussion and Conclusion}

In this paper, we study the ability of molecular rotors to characterize the intracellular rheology of red blood cells. We identified a molecular rotor (DASPI: trans-4-[4-(dimethylamino)-styryl]-1-methylpyridiniiodide), that is suitable for the study of hemoglobin and intracellular medium of red blood cells. For this purpose, the excitation-emission spectra of DASPI were measured in selected solutions: DMSO, ethylene glycol, glycerol and hemoglobin. In all fluids, DASPI presented a fluorescence excitation peak around $480$ nm and emission peak around $600$ nm. We have thus shown that the excitation-emission spectrum of DASPI does not overlap with the intrinsic fluorescent spectrum of hemoglobin in the ultraviolet domain \cite{Waterman1978}. Moreover, its emission spectrum is in the range where the absorbance of hemoglobin is lowest ($\sim 600$ nm, Fig.~\ref{fig:abs}), which allows it to be used at moderate concentrations.

Tested in simple solutions such as ethylene glycol/glycerol and glycerol/water solutions, DASPI shows sensitivity to the local viscosity $\eta$, and exhibits a Förster-Hoffmann dependency in a range covering two decades, $\log \phi= C + b \log \eta$, with $\phi$ the LE-state quantum yield and $b$ and $C$ two solvent and dye-dependent parameters. The exponent $b$, which is a measure of the efficiency of the molecular rotor in the solvent, was found to be around $0.7$ in both solutions, which is close to the predicted theoretical values of $2/3$\cite{forster1971} or $0.6$\cite{loufty1982}. The parameter $C$ increases by about one and a half decade when the concentration of DASPI increases by two decades, which allows to adjust the dye concentration to the sensitivity of the intensity reading device in potential applications. 
The comparison between ethylene glycol/glycerol and glycerol/water solutions -- one where the polarity effects are minimized, the other with variable polarity -- confirms the previous result obtained by Haidekker and
colleagues\cite{haidekker2005} that there is no impact of variable polarity on obtaining a Förster-Hoffmann equation between quantum yield and viscosity. In addition, our data show that temperature has little influence on the $C$ parameter, which allows us to use it as a RBC stiffening control parameter in the section \ref{sec:stiffening}.

The penetration of the molecular rotor (DASPI) has been studied in healthy red blood cells. We showed 
using confocal microscopy that the molecular rotor spontaneously penetrates into red blood cells. With a depth of field of approximately $600$ nm, confocal microscopy allows us to unambiguously separate the fluorescence response of the cytosol and the lower and upper membranes as soon as the focal plane is far enough away from the membrane. 

We therefore revealed that the molecular rotor is uniformly distributed throughout the volume of RBCs, giving evidence that it is a suitable probe for characterizing their intracellular rheology. As fluorescence profiles include a contribution from the membrane with a locally strong signal, the specific interactions between DASPI and the cell membrane should receive particular attention in future studies in order to reach quantitative understanding of the influence of structural and rheological properties of membranes on MR fluorescence.

In order to highlight and characterize the sensitivity of DASPI to RBC rheological properties, we varied temperature as a way to control cytosol and membrane properties, as a decrease in temperature leads to an overall stiffening of RBCs. Results show that for a given healthy blood sample, the measured fluorescence intensity at the whole cell level varies nearly by a factor 2 between $18^\circ$C and $43^\circ$C in correlation with variations of cytosol viscosity and membrane elasticity. We showed that by making reasonable assumptions about the RBC geometry, it is possible to separate and quantify the respective contributions of the cytosol and the membrane to the overall fluorescence signal and thus to derive information on the modifications of both components due to temperature change. 
We show in particular that the estimated contribution from the cytosol reproduces well the intrinsic heterogeneity of the RBC sample, and is consistent with independent measurements on hemoglobin solutions.
Finally, by simply varying temperature for healthy cells, we demonstrate by this example that in principle the DASPI molecular rotor can be used to derive measurements of intracellular viscosity and quantify sample heterogeneity.

For a more quantitative analysis, the partial absorption of both incident and emitted light by hemoglobin will have to be modeled and taken into account, which offers prospects for future studies. Further investigations should also focus on calibration curves of the DASPI response in hemoglobin solutions, taking into account hemoglobin absorbance. While our study shows that the rotor response is unequivocally correlated to rheological changes in healthy hemoglobin induced by temperature variations, these results should be generalized to systematic variations in hemoglobin concentration, as well as to studies on pathological situations where the structure of hemoglobin may be altered.

Our results are a proof of concept that MRs can be used as nanoscale rheological probes in small compartments such as red blood cells. Their use in this context of evaluation of mechanical properties at the cellular level is a promising prospect, in particular because it would make it possible to measure the distribution of cell properties over large samples, unlike other techniques that only allow the measurement of average values over the entire sample.
The alteration and dispersion of cell properties in pathological cases can indeed be critical for clinical aspects and such a technique opens new perspectives for medical diagnosis purposes by providing more detailed and quantitative information through a rather simple and direct measurement of fluorescence.

\section*{Conflicts of interest}
There are no conflicts to declare.

\section*{Acknowledgements}
The authors wish to thank C. Makowski and B. Polack from CHU Grenoble-Alpes for helpful discussions on blood and hemoglobin preparation. F. Molina (Sys2Diag Montpellier) and A. Pasturel (MSC Paris \& Sys2Diag Montpellier) are acknowledged for fruitful discussions.
The authors thank O. Stefan (LIPhy) for help with hemoglobin purification. T. P. acknowledges support of CNES for investigations on blood rheology and flows. 

\bibliography{biblio}
\bibliographystyle{amsplain}

\end{document}